\documentclass[12pt,a4paper]{article}
\usepackage{a4wide}
\usepackage{graphicx}
\usepackage{epsfig}
\usepackage{amsfonts}

\def    \beq           	{\begin{equation}}
\def    \eeq           	{\end{equation}}
\def    \bea           	{\begin{eqnarray}}
\def    \eea           	{\end{eqnarray}}
\def    \nn            	{\nonumber}  
\def    \raw           	{\rightarrow}

\def	\cN		{{\mathcal{N}}}

\def	\cO		{{\mathcal{O}}}
\def	\cW		{{\mathcal{W}}}
\def 	\bZ		{{\mathbb Z}}

\newcommand{\id}{1\!\!1}

\renewcommand{\thefootnote}{\fnsymbol{footnote}}

\begin{document}

\title{
\begin{flushright}
\ \\*[-80pt]
\begin{minipage}{0.2\linewidth}
\normalsize
KUNS-2060 \\*[50pt]
\end{minipage}
\end{flushright}
{\Large \bf
Heterotic orbifold models on Lie lattice \\
with discrete torsion
\\*[20pt]}}

\author{
Kei-Jiro Takahashi\footnote{
E-mail address: keijiro@gauge.scphys.kyoto-u.ac.jp}\\*[20pt]
{\it \normalsize
Department of Physics, Kyoto University,
Kyoto 606-8502, Japan} \\
 \\*[50pt]}

\date{
\centerline{\small \bf Abstract}
\begin{minipage}{0.9\linewidth}
\medskip
\medskip
\small
We provide a new class of ${\mathbb Z}_N \times {\mathbb Z}_M$ heterotic orbifolds on non-factorizable tori, whose boundary conditions are defined by Lie lattices. Generally, point groups of these orbifolds are generated by Weyl reflections and outer automorphisms of the lattices. We classify abelian orbifolds with and without discrete torsion. Then we find that some of these models have smaller Euler numbers than those of models on factorizable tori $T^2 \times T^2 \times T^2$. There is a possibility that these orbifolds provide smaller generation numbers of $\cN=1$ chiral matter fields than factorizable models.
\end{minipage}
}

\begin{titlepage}
\maketitle
\thispagestyle{empty}
\end{titlepage}

\renewcommand{\thefootnote}{\arabic{footnote}}
\setcounter{footnote}{0}

\section{Introduction}
\hspace{.19in}
The standard model is a remarkably successful theory 
which explains experimental data with good accuracy. 
It is a chiral gauge theory 
with the gauge group SU(3)$\times$SU(2)$\times$U(1)
and three generations of quarks and leptons.
On the other hand, 
we can not answer why there exist three generations of matter, 
and dark matter in the universe. 
It is a challenging issue to elucidate 
the origin of parameters of the standard model, 
and to understand the deep structure of nature. 
Superstring theory is a most promising candidate to give the explanation. 
Heterotic orbifold constructions are interesting attempts to 
realize realistic string models \cite{Dixon:1985jw, Dixon:1986jc}.
Some of $\cN$=1 supersymmetric models include three generations of 
chiral fields with SU(3)$\times$SU(2)$\times$U(1)${^n}$. 
In spite of such a nice feature, 
it seems difficult to derive realistic Yukawa matrices. 
So as to reproduce the complete spectrum, 
it is important to study new types of compactifications. 

Most of heterotic orbifold models constructed so far are 
based on abelian discrete groups ${\mathbb Z}_N$ and 
${\mathbb Z}_N \times {\mathbb Z}_M$ 
\cite{Kobayashi:2004ud,Forste:2004ie,Forste:2005rs}. 
These abelian orbifold models are classified 
by Coxeter elements, 
which are subgroups of Weyl group of Lie lattice \cite{Bailin:1999nk}.
In ${\mathbb Z}_N \times {\mathbb Z}_M$ Coxeter orbifold models 
the compact spaces are factorized to 
$T^2 \times T^2 \times T^2$.
However the geometry of compact space would be non-factorizable in general, 
just like Calabi-Yau threefold. 
Recently non-factorizable ${\mathbb Z}_2 \times {\mathbb Z}_2$ 
orbifold models of E${}_8\times$E${}_8$ heterotic string 
are examined \cite{Faraggi:2006bs}
\footnote{In context of crystallographical symmetry 
some non-factorizable models are presented in \cite{Dixon:1986jc}.}.
Due to the tilted structure of compact space 
the fixed tori by $\bZ_2$ action have nontrivial geometry, 
and the number of fixed tori can be less than that of factorizable models.
In some cases 
the structures of Yukawa interactions are also changed \cite{Forste:2006wq}.

We find that
${\mathbb Z}_2 \times {\mathbb Z}_3$, 
${\mathbb Z}_2 \times {\mathbb Z}_4$ 
and 
${\mathbb Z}_4 \times {\mathbb Z}_4$ non-factorizable orbifolds are
allowed by automorphisms of Lie lattice 
and have $\cN=1$ chiral massless modes. 
${\mathbb Z}_N \times {\mathbb Z}_M$ orbifold models 
with $N,M \geq 2$ allow the addition of discrete torsion \cite{Vafa:1986wx}, 
and the generation numbers of matter can be significantly reduced 
\cite{Font:1988mk, Kobayashi:1990fx}.
We classify these new abelian orbifold models in the standard embedding 
of gauge group with and without discrete torsion, 
and find some of these models have different Euler numbers 
from the factorizable case.

The paper is organized as follows. 
In section 2 we review the heterotic orbifold, 
and some conditions for model buildings.
In section 3 we give an example of 
${\mathbb Z}_2 \times {\mathbb Z}_2$ orbifold model, 
and explain the idea of non-factorizable orbifold.
Section 4 addresses model building of 
${\mathbb Z}_N \times {\mathbb Z}_M$ orbifold.
We give some examples and detail calculation of these models. 
Section 5 is devoted to conclusion and discussions.
All results of ${\mathbb Z}_N \times {\mathbb Z}_M$ models 
on non-factorizable tori are listed in appendix A, 
and tables would be convenient for future works.
Appendix B contains the basics of Weyl reflection, Coxeter elements 
and their explicit representations. 
By the use of these representations, 
we construct all point group elements of orbifold in our paper.

\section{Automorphism of Lie lattice}

\subsection{Weyl reflection and orbifold}
\hspace{.19in}
We compactify six dimensions on a torus $T^6$ 
in order to construct four dimensional models.
$T^6$ is obtained by compactifying ${\mathbb R}^6$ 
on a lattice $\Lambda$, 
\beq
T^6 = R^6/\Lambda.
\eeq
Points in ${\mathbb R}^6$ differing by a lattice vector $L \in \Lambda$ 
are identified as follows, 
\beq
x \sim x + 2\pi L.
\eeq
In heterotic string construction, 
the resulting  spectrum has $\cN=4$ supersymmetry.
This theory is non-chiral, and 
there is no matter field with bi-fundamental representation.
It is interesting  
to consider orbifold \cite{Dixon:1985jw, Dixon:1986jc} 
in order to obtain a chiral theory with $\cN=1$ supersymmetry. 
An orbifold is defined to be the quotient of a torus 
over a discrete set of isometries of the torus, 
called the point group $P$, i.e.
\beq
\cO = T^6/P=R^6/S. 
\eeq
Here $S$ is called the space group, and is the semidirect product of the 
point group $P$ and the translation group. 
In this paper we consider the case 
that $\Lambda$ is generated by Lie lattice.
For any simply laced Lie group of rank $l$ and dimension $d$, 
the Lie lattice is given as 
\beq
\Lambda \equiv \{ \sum_{i=1}^l n_i \alpha_i | n_i \in \bZ \}.
\eeq

The point group of orbifold must be automorphisms of the lattice.
The automorphisms of the Lie lattice can be realized by 
Weyl reflection and outer automorphism 
which includes graph automorphism of Dynkin diagram.
The Weyl group $\cW$ is generated by the following Weyl reflections
\beq
r_{\alpha_k}:\lambda \raw \lambda -2{\alpha_k \cdot \lambda \over \alpha_k \cdot \alpha_k} \alpha_k. 
\eeq
The point group, $\theta$ and $\phi$, 
of ${\mathbb Z}_N \times {\mathbb Z}_M$ orbifold 
on Lie lattice can be defined by 
two commutative elements from Weyl group and outer automorphism 
$G_{out}$, i.e.
\beq
\theta,~\phi \in \{ \cW,G_{out} \}.
\eeq
Here, Weyl reflections $r_{\alpha} \in \cW$ 
and outer automorphisms $g \in G_{out}$ generate larger group than $\cW$, 
then we write this group as $\{ \cW,G_{out} \}$.
If we select pairs of elements $\theta$, $\phi$ 
as the generators of point groups, 
we can construct hundreds of 
${\mathbb Z}_N \times {\mathbb Z}_M$ orbifold models\footnote{
See Appendix.B for explicit realization of these elements.}. 
However in ${\mathbb Z}_2 \times {\mathbb Z}_2$ models 
there are only 8 classes \cite{Forste:2006wq}, 
which have different numbers of twisted sectors. 
This is because the six dimensional tori 
in ${\mathbb Z}_2 \times {\mathbb Z}_2$ models are decomposed to
two or three dimensional tori 
by continuous deformation of geometric moduli, 
and the numbers of 
fixed points or fixed tori by 
${\mathbb Z}_2$ action on these decomposed torus is 1,2, or 4.
The other ${\mathbb Z}_N \times {\mathbb Z}_M$ orbifold models 
are also classified to several classes as we will see. 

\begin{figure}[tbp]
\begin{center}
\includegraphics{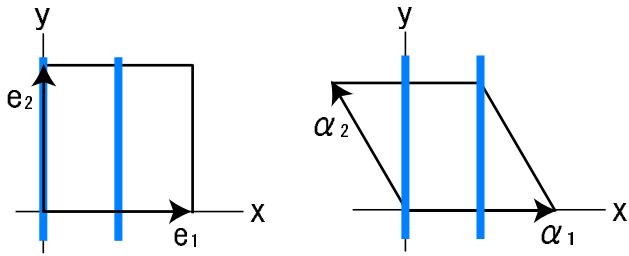}
\caption{${\mathbb Z}_2 \times {\mathbb Z}_2$ orbifold.
The left figure is factorizable orbifold, 
and the right one is non-factorizable one.
Blue colored lines are fixed tori $T^1$.}
\label{so(4)su(3)}
\end{center}
\end{figure}

The idea of non-factorizable orbifold can be understood easily 
by two dimensional example. 
The Cartesian coordinates $e^1$ and $e^2$ are given by
\bea
e^1 &=& (1,0),\nn \\
e^2 &=& (0,1).
\eea
We call this basis as the SO(4) root lattice. 
The basis of SU(3) root lattice is given by simple roots,
\bea
\alpha^1 =& (1,0),\nn \\
\alpha^2 =& \left( -\frac12,\frac{\sqrt{3}}{2} \right).
\eea
We choose the orbifold action as,
\beq
\theta \equiv r_{e_1}= r_{\alpha_1} =
\left(
\begin{array}{cc}
-1 & 0 \\
0 & 1
\end{array}
\right). 
\eeq
When $\theta$ acts on the SO(4) root lattice, 
there are two fixed tori and 
$\theta$-twisted sector strings live on these tori. 
On the other hand, $\theta$ action on SU(3) root lattice generates only 
one fixed torus, 
because two lines in figure \ref{so(4)su(3)} are 
connected owing to the tilted structure of lattice.
The SU(3) model has only a half number of 
twisted sectors than that of SO(4) model, 
and is a simple example of non-factorizable orbifold.
This mechanism can be applied to higher dimensional Lie lattices, 
and generally the torus can not be factorized 
to $T^2 \times T^2 \times T^2$.
We classify all ${\mathbb Z}_N \times {\mathbb Z}_M$ orbifold 
on SU($N$) and SO($N$) Lie lattice\footnote{
As we include Weyl group and graph automorphisms to point group, 
SU(3) and G${}_2$ root lattices are equivalent, and 
the same for SO(8) and F${}_4$, SO($2N$) and Sp($2N$),
SO($2N+1$) and SU($2$)${}^N$ respectively. 
Of course these equivalences are limited to the lattice structures $\Lambda$, 
not the roots.}.

\subsection{Supersymmetric condition}
\hspace{.19in}
One can always choose diagonal basis for $\theta$ on SO(6): 
\beq
\theta = \exp(2\pi i (a J_{12}+ b J_{34}+ c J_{56})).
\label{eigenstate}
\eeq
Then the eight eigenvalues of $\theta$ acting 
on the spinor of SO(8) are $e^{\pi i (\pm a + \pm b +\pm c)}$.
To preserve $\cN=1$ supersymmetry at least two 
of them should be left invariant. 
For convenience we list the thirteen shift vectors $(|a|,|b|,|c|)$ 
which are allowed by these conditions \cite{Dixon:1986jc}.
\begin{table}[h!]
\renewcommand{\arraystretch}{1.5}
\normalsize
\begin{center}
\begin{tabular}{ccccc}
\hline
 & $(|a|,|b|,|c|)$ & \\
\hline
\hline
 $(0, \frac12, \frac12)$ & $(0, \frac13, \frac13)$ & $(\frac13, \frac13, \frac23)$  & $(0, \frac14, \frac14)$ & $(\frac14, \frac14, \frac12)$\\
 $(0, \frac16, \frac16)$ & $(\frac16, \frac16, \frac13)$ & $(\frac16, \frac13, \frac12)$  & $(\frac17, \frac27, \frac37)$ & $(\frac18, \frac14, \frac38)$\\
 $(\frac18, \frac38, \frac12)$ & $(\frac{1}{12}, \frac13, \frac{5}{12})$ & $(\frac{1}{12}, \frac{5}{12}, \frac12)$  &   &  \\
\hline
\end{tabular}
\end{center}
\caption{The shift vectors of point group $\theta$, 
which satisfy the tachyon free condition.}
\label{tachyon-free}
\end{table}

In this paper 
${\mathbb Z}_2 \times {\mathbb Z}_2$, 
${\mathbb Z}_2 \times {\mathbb Z}_3$, 
${\mathbb Z}_2 \times {\mathbb Z}_4$ 
and 
${\mathbb Z}_4 \times {\mathbb Z}_4$ models 
are investigated. 
In following models, 
we can always select 
the shift vectors $v$ ($w$) of point group elements $\theta$ ($\phi$) of 
${\mathbb Z}_N \times {\mathbb Z}_M$ as
\bea
\theta :~ v &=& (n,-n,0) \nn \\
\phi :~ w &=& (0,m,-m) 
\eea
where $n,m= \frac{1}{2}$ or $\frac{1}{3}$ or $\frac{1}{4}$. 
This projects out six components of the spinor, 
and leaves two chiral spinors 
$|\pm (\frac12,  \frac12, \frac12, \frac12) \rangle$ invariant.
Then $\cN=1$ supersymmetry will be unbroken in four dimension.

A point group element of ${\mathbb Z}_N$ orbifold, 
which has no eigenvalues equal to 1 
when acting on the six dimensional Lie lattice, 
is called non-degenerate. 
These elements which are generated by Weyl reflection 
on six dimensional Lie lattice, 
are classified by Carter diagram in ref.\cite{Schellekens:1987ij}.
In Appendix.B we review Carter diagram and Coxeter elements shortly.

\subsection{Modular invariance and discrete torsion}
\hspace{.19in}
Heterotic orbifold models must satisfy some consistency conditions 
required by the modular invariance. 
The modular invariance guarantees the anomaly cancellation 
in low energy theory \cite{Vafa:1986wx}.
For $Z_N$ orbifolds with prime $N$, 
the level matching conditions are 
necessary and sufficient for modular invariance to all loops 
of string amplitude.
For the $\theta^k \phi^l$-twisted sector, 
the level matching condition is 
\bea
& N^\prime [(kV+lW)^2 -(kv+lw)^2] = 0 \bmod2 , \\
& k=0,\cdots,N-1,~~l=0,\cdots,M-1, \nn
\eea
where $N^\prime$ is the order of the twist $\theta^k \phi^l$, 
and $V$ and $W$ are 
the shift vectors in gauge sector 
associated to $\theta$ and $\phi$ respectively. 
In our paper we consider E${}_8\times$E${}_8^\prime$ heterotic string 
models with standard embedding. 
The shift vectors of gauge sector E${}_8$ are given as 
\bea
v= (n, -n, 0) ~&\raw& V= (n,-n,0,0,0,0,0,0), \\
w= (0,m, -m) ~&\raw& W= (0,m,-m,0,0,0,0,0). \nn
\eea 
Thus the level matching condition is trivially satisfied 
in the standard embedding. 
This corresponds to embedding the spin connection in 
the gauge connection. 

Turning on background antisymmetric field $B_{\mu\nu}$ on the torus
it introduces phases to string state \cite{Vafa:1986wx}\cite{Font:1988mk}. 
This effect can be described by
the general form of one-loop partition function,
\beq
Z= \frac{1}{MN} \sum_{\theta, \phi} \epsilon(\theta, \phi) Z(\theta, \phi).
\eeq
The phase $\epsilon(\theta, \phi)$ is called discrete torsion.
In a ${\mathbb Z}_N$ orbifold 
these phase are fixed by one-loop modular invariance.
On the other hand 
in ${\mathbb Z}_N \times {\mathbb Z}_M$ orbifolds,
where $M$ is generally divisible by $N$, 
the phase is restricted to $N$-th root of unity,
\beq
\epsilon(\theta, \phi)\equiv \epsilon,~ \epsilon^N=1. 
\eeq
Then the phases for general twisted sectors are given by
\beq
\epsilon(\theta^k \phi^l, \theta^t \omega^s) =
\epsilon^{(ks-lt)}.
\eeq

For orbifolds with non-prime $N$, 
we need to generalize the GSO projection 
\cite{Font:1988mk}.
The number of $\theta^k \phi^l$-twisted states is given by 
\beq
D(\theta^k \phi^l) = \frac{1}{MN} 
\sum_{t=0}^{N-1} \sum_{s=0}^{M-1} 
\epsilon^{(ks-lt)} 
\tilde{\chi}(\theta^k \phi^l, \theta^t \omega^s) 
\Delta(k,l; t,s),
\label{twist-number}
\eeq
where $\tilde{\chi}$ is the number of points left simultaneously fixed 
by $\theta^k \phi^l$ and $\theta^t \phi^s$. 
If $\theta^k \phi^l$ leaves unrotated some of the coordinates, 
$\tilde{\chi}$ must be calculated 
using only the sub-lattice which is rotated.
$\Delta(k,l; t,s)$ is a state dependent phase.
In the case of standard embedding it is given by 
\beq
\Delta(k,l; t,s) = P^{(k,l)} \exp\{2\pi i 
[(p+kV+lW)(tV+sW) -(q+ kv+lw)(tv+sw)]\}, 
\label{phase-orbi}
\eeq
where $P^{(k,l)}$ indicates a contribution of oscillators. 
$p$ is momentum of the E${}_8\times$E${}_8^\prime$ gauge sectors, 
and $q$ is H-momentum of the twisted states.

The generalization of the Euler characteristic of $Z_N$ orbifold is given by 
\beq
\chi_{\epsilon}
= \frac{1}{MN} \sum_{k,t=0}^{N-1} \sum_{l,s=0}^{M-1} 
\epsilon^{(ks-lt)} 
\chi(\theta^k \phi^l, \theta^t \omega^s), 
\label{euler-disc}
\eeq
where $\chi$ is the number of points left simultaneously fixed 
by $\theta^k \phi^l$ and $\theta^t \phi^s$.
The number of generations is equal to $\chi_{\epsilon}/2$.
In section 4, we construct orbifold models 
for allowed values of $\epsilon$.

\section{${\mathbb Z}_2 \times {\mathbb Z}_2$ orbifold: Factorisable VS non-factorizable}
\hspace{.19in}
${\mathbb Z}_2 \times {\mathbb Z}_2$ orbifold 
is phenomenologically interesting model, 
because some three generation models are presented 
with the aid of Wilson lines, 
and three generations may be associated to 
three complex dimension of compact space \cite{Forste:2004ie}.

In heterotic orbifold model 
there are two classes of string states.
One is untwisted sector in bulk 
and the other is twisted sector 
which localizes at fixed torus.
The chirality of untwisted sector of 
${\mathbb Z}_2 \times {\mathbb Z}_2$ is left-right symmetric, 
so it does not contribute to the number of generations. 
The number of zero modes of twisted sector is related to
the number of fixed tori.
In a factorizable model with the action $\theta$ and $\phi$, 
whose shift vectors are 
$v=(\frac12,-\frac12,0)$ and $w=(0,\frac12,-\frac12)$ respectively, 
the number of tori of $\theta$-twisted sector is $4^2$,
because there are four fixed points in the first and second tori, 
and the third torus is free from the action of $\theta$.
Therefore the total number of zero modes of three twisted sectors is 48, 
and this corresponds to the generation numbers of $\cN=1$ chiral matter 
which have gauge charge $27 \in$ E${}_6$ 
in the standard embedding. 

We can confirm this result by calculating the Euler number, 
because the number of generations 
is equal to a half of Euler number \cite{Dixon:1985jw},
\beq
\chi= \frac{1}{N} \sum_{[g,h]=0} \chi_{g,h},
\label{euler}
\eeq
where $\chi_{g,h}$ is the number of fixed points 
by action $g$ and $h$, and $N$ is the order of point group.
For ${\mathbb Z}_2 \times {\mathbb Z}_2$ orbifold, 
this equation is simplified to
\beq
\chi= \frac{3}{2} \chi_{\theta,\phi}.
\eeq
Here, $\chi_{\theta,\phi}$ is the number of points left 
simultaneously fixed by $\theta$ and $\phi$, and
is equal to $4^3$.
Then we have $\chi=96$, 
and this agrees with the former result.

In the case of non-factorizable model 
it is easier to use Lefschetz fixed point theorem \cite{Faraggi:2006bs}.
The number of fixed tori (\#FT) of $\theta$-twisted sector is
\beq
\#FT={vol((1-\theta)\Lambda) \over vol(N)},
\eeq
where $N$ is the lattice normal to the sub-lattice invariant by the action.
%
%

%
\begin{figure}[tb]
\begin{center}
\includegraphics{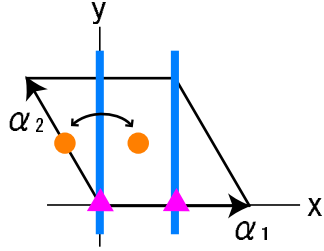}
\caption{${\mathbb Z}_2 \times {\mathbb Z}_2$ orbifold.
The blue colored lines are fixed tori by action of $\theta$.
The circles and triangles are fixed points by action of $\phi$.
The twisted states on the circles are mapped to the other by $\theta$, 
and the linear combinations of states are eigenstates of orbifold.}
\label{su(3)fix}
\end{center}
\end{figure}
As an example we consider 
the orbifold model on SU(3)$\times$SO(8) root lattice,
whose basis is given by simple roots,
\bea
\alpha_{1} &=& \sqrt{2}(1,0,0,0,0,0),\nn \\
\alpha_{2} &=& \sqrt{2}\left(-\frac12,\frac{\sqrt{3}}{2},0,0,0,0\right),\nn \\
\alpha_{3} &=&(0,0,1,-1,0,0),\nn \\
\alpha_{4} &=&(0,0,0,1,-1,0),\\
\alpha_{5} &=&(0,0,0,0,1,-1),\nn \\
\alpha_{6} &=&(0,0,0,0,1,1),\nn 
\eea
and 
${\mathbb Z}_2 \times {\mathbb Z}_2$ orbifold actions, 
$\theta$, $\phi$, are given by 
\bea
\theta 
= \left(\begin{array}{cccccc}
-1 & 0 & 0 & 0 & 0 &0 \\
0 & 1 & 0 & 0 & 0 &0 \\
0 & 0 & 1 & 0 & 0 &0 \\
0 & 0 & 0 & -1 & 0 &0 \\
0 & 0 & 0 & 0 & -1 &0 \\
0 & 0 & 0 & 0  & 0 &-1 
\end{array} \right),\nn \\
\phi 
= \left(\begin{array}{cccccc}
-1 & 0 & 0 & 0 & 0 &0 \\
0 & -1 & 0 & 0 & 0 &0 \\
0 & 0 & -1 & 0 & 0 &0 \\
0 & 0 & 0 & -1 & 0 &0 \\
0 & 0 & 0 & 0 & 1 &0 \\
0 & 0 & 0 & 0  & 0 &1 
\end{array} \right).
\eea
The common fixed points by the actions of 
$\theta$ and $\phi$ are,
\bea
&(0,0,0,0,0,0),~(0,0,1,0,0,0),~(0,0,0,0,\frac12,\pm\frac12),\nn \\
&(\frac{\sqrt{2}}{2},0,0,0,0,0),~(\frac{\sqrt{2}}{2},0,1,0,0,0),
~(\frac{\sqrt{2}}{2},0,0,0,\frac12,\pm\frac12),\nn \\
&~(0,0,\frac12,\frac12,
\underline{\pm \frac12,0}),~(\frac{\sqrt{2}}{2},0,\frac12,\frac12,
\underline{\pm \frac12,0}),
\eea
where the underlined entries can be permuted.
This leads to $\chi_{\theta,\phi}=16$ and $\chi=24$.
Then the generation number is twelve.
We can reconfirm this result by counting 
the number of fixed tori as follows.
There are four independent fixed tori of the $\theta$-twisted sector,
\bea
&(0,x,y,0,0,0),\nn \\
&(0,x,y,\underline{\frac12,\frac12,0}).
\eea
Note that these tori are identified 
by the sub lattice $(1-\theta)\Lambda$.
The $\theta \phi$-twisted sector also has four fixed tori.
In the $\phi$-twisted sector there are eight fixed tori,
\bea
&(0,0,0,0,x,y),~(0,0,\frac12,\frac12,x,y),
(\frac{\sqrt{2}}{2},0,0,0,x,y),~
(\frac{\sqrt{2}}{2},0,\frac12,\frac12,x,y),\nn \\
&(\pm \frac{\sqrt{2}}{4},\frac12 \sqrt{\frac32},0,0,x,y),
~(\pm \frac{\sqrt{2}}{4},\frac12 \sqrt{\frac32},\frac12,\frac12,x,y).
\eea
Then the total number of tori is 16,
and does not match the generation number.

This is because two of fixed tori of the $\phi$-twisted sector 
are not invariant by the action of $\theta$. 
On the SU(3) torus in figure \ref{su(3)fix}, 
there are four fixed points, 
and  can be labeled by shift vector, 
$v= n \alpha_1 + m \alpha_2,~ n,m=0,1$.
The $\theta$-invariant states are
\beq
|0 \rangle,~~|\alpha_1 \rangle.
\eeq
These are charged matter of representation $27$.
We can take linear combination of remaining two states 
as eigenstates of action of $\theta$,
\bea
+1:&~|\alpha_2\rangle + |\alpha_1+\alpha_2 \rangle,\nn \\
-1:&~|\alpha_2\rangle - |\alpha_1+\alpha_2 \rangle, 
\eea
where $\pm1$ denote the eigenvalue of these states 
under the action of $\theta$. 
The phase of physical states should be cancelled 
with $\Delta$-phase (\ref{phase-orbi}). 
These are the same chirality states 
with the charge of representation in $27$ and $\overline{27}$ respectively, 
and they do not contribute to the number of generations \cite{Faraggi:2006bs}.
The generation number from the $\phi$-sector is four, 
and we have twelve generations from three twisted sectors,
which is equal to a half of $\chi$.
This is significantly small compared to 
the generation number of factorizable model.

\begin{figure}[tbp]
\begin{center}
\includegraphics{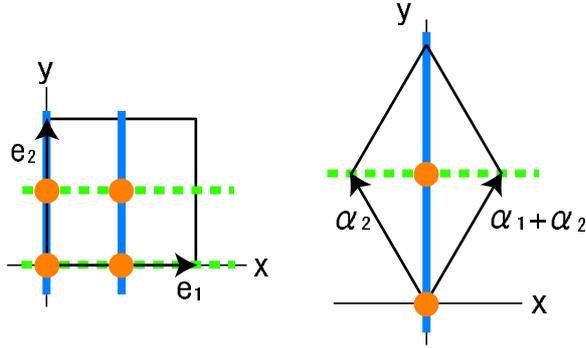}
\caption{${\mathbb Z}_2 \times {\mathbb Z}_2$ orbifold.
Blue colored lines are fixed tori by $\theta$ action, 
and green colored dotted lines are by $\theta \phi$ action.
The circles are common fixed points.}
\label{so(4)su(3)II}
\end{center}
\end{figure}
The diminution of fixed tori and fixed points in SU(3) torus
can be seen in figure \ref{so(4)su(3)II}. 
In this figure the basis of the SU(3) root lattice is changed to
$\alpha_1+\alpha_2$ and $\alpha_1$, 
and that makes it easier to draw the orbifold action.
In this figure we can see 
that the decrease of $\theta$ fixed tori of the non-factorizable model
is related to direction which is left invariant by the action of $\theta$. 
Therefore the diminution does not occur 
in non-degenerate orbifold such as Coxeter orbifold 
which rotates the whole space of $T^6$.

\section{${\mathbb Z}_N \times {\mathbb Z}_M$ non-factorizable orbifold}
\hspace{.19in}
As we have seen in ${\mathbb Z}_2 \times {\mathbb Z}_2$ model, 
the number of fixed tori on non-factorizable orbifold 
is less than that of factorizable model 
due to the topological difference of fixed tori.
The following shift vectors of point group elements 
which generate tori are
\bea
&(0,\frac12,\frac12),~(0,\frac13,\frac13),~
(0,\frac14,\frac14),~(0,\frac16,\frac16).&
\eea
We call these elements degenerate. 
That means some eigenvalues of 
the point group element are equal to one. 
Then the model has $\cN=2$ supersymmetry and non-chiral.
In order for the resulting orbifold model to have $\cN=1$ supersymmetry 
the point group of orbifold should be
${\mathbb Z}_N \times {\mathbb Z}_M$, 
and this is totally non-degenerate.
We can embed some of these point group elements 
to non-factorizable tori.

Non-factorizable tori are defined by root lattices.
Lie lattice of SO($2N$) are given by the simple roots, 
\bea
\label{so(2n)root} 
\alpha_{1} &=&(1,-1,0,0,\cdots,0),\nn \\
\alpha_{2} &=&(0,1,-1,0,\cdots,0),\nn \\
& \vdots & \nn \\
\alpha_{N-1} &=&(0,\cdots,0,1,-1),\\
\alpha_{N} &=&(0,\cdots,0,1,1),\nn
\eea
and simple roots of SU($N$) are 
\bea
\alpha_{1} &=&(1,-1,0,0,\cdots,0),\nn \\
\alpha_{2} &=&(0,1,-1,0,\cdots,0),\nn \\
& \vdots & \nn \\
\alpha_{N-1} &=&(0,\cdots,1,-1,0),\\
\alpha_{N} &=&(0,\cdots,0,1,-1).\nn 
\label{su(n)root} 
\eea
For simplicity we use the basis 
for SU(2) root lattice,
\beq
\alpha_{1}=(1).
\label{su(2)root} 
\eeq
Hereafter we use direct sum of these simple roots 
for the basis of compact space $T^6$.

\subsection{${\mathbb Z}_2 \times {\mathbb Z}_4$ models}
\hspace{.19in}
As an example we consider 
a model on SO(12) root lattice, 
which is the case that $N=6$ in (\ref{so(2n)root}).
The only consistent point group action on this lattice, 
except its conjugate representation, is
\bea
\theta&=&\left(
\begin{array}{ccccccc}
0 & -1 & 0 & 0 & 0 & 0 \\
1 & 0 & 0 & 0 & 0 & 0 \\
0 & 0 & 0 & 1 & 0 & 0 \\
0 & 0 & -1 & 0 & 0 & 0 \\
0 & 0 & 0 & 0 & 1 & 0 \\
0 & 0 & 0 & 0 & 0 & 1 
\end{array}
\right),\\
\phi&=&
diag(1,1,-1,-1,-1,-1).\nn 
\eea

We count the number of states 
with representations $27$ and $\overline{27}$ 
by the use of the coordinates of fixed points and fixed tori 
\footnote{In this approach we can observe the twisted states explicitly. 
However we can systematically count these numbers 
by (\ref{twist-number}).}.
The $\theta^i \phi^j$-twisted sector $T_{ij}$ localizes 
at fixed points or tori as follows, 
\bea
\begin{array}{ll}
T_{01}: & (x,y,0,0,0,0),~(x,y,\underline{\frac12, \frac12,0,0}),~
(x,y,\frac12, \frac12,\frac12, \frac12), \\
T_{10}: & (0,0,x,y,0,0),~(\frac12, \frac12,x,y,\frac12, \frac12), \\
T_{11}: & (0,0,0,0,0,0),~(1,0,0,0,0,0),~
(\frac12, \frac12,0,0,\frac12, \pm \frac12), \\
 & (\frac12, \frac12,\underline{\pm \frac12, 0},0,0),~
(0,0,\underline{\frac12, 0},\frac12, \pm \frac12), \\
 & (0,0,\frac12, \pm \frac12,0,0),~
(\frac12, \frac12,\frac12, \frac12,\frac12, \pm \frac12), \\
T_{20}:&(0,0,x,y,0,0),~(\frac12,\frac12,x,y,0,0),~(0,0,x,y,\frac12,\frac12) \\
  & (\underline{\frac12,0},x,y,\underline{\frac12,0}),~
(\frac12, \frac12,x,y,\frac12, \frac12), \\
T_{21}: & (0,0,0,0,x,y),~(\underline{\frac12, \frac12,0,0},x,y),~
(\frac12, \frac12,\frac12, \frac12,x,y).
\end{array}
\eea

In the orbifolds of ${\mathbb Z}_N$ with non-prime $N$, 
the physical state of $\theta^k$-sector is generally linear combination 
of state at fixed points by the action $\theta^k$ 
\cite{Kobayashi:1990fx, Bailin:1999nk}.
If $f_k$ is a fixed point of $\theta^k$ 
such that $l$ is the smallest number giving 
$\theta^l f_k = f_k +u,~u\in \Lambda$, 
then the eigenstates of $\theta$ are 
\beq
|p \rangle = \sum_{r=0}^{n-1} e^{i \gamma r} |\theta^r f_k \rangle, 
\eeq
with $\gamma=2 \pi /l$, $p=0,1,\cdots,N-1$.
Then the physical states of $T_{01}$ sector by orbifold are 
linear combinations of them, 
\bea
1:& |(x,y,0,0,0,0)\rangle,~
|(x,y,\frac12, \frac12,\frac12, \frac12)\rangle,\\
& |(x,y,\frac12, \frac12,0, 0)\rangle,~
|(x,y,0,0,\frac12, \frac12)\rangle, \nn\\
& |(x,y,\frac12, 0,\frac12, 0)\rangle + 
|(x,y,0, \frac12,\frac12, 0)\rangle, \nn\\
& |(x,y,\frac12, 0, 0,\frac12)\rangle + 
|(x,y,0, \frac12, 0,\frac12)\rangle, \nn\\
-1:& |(x,y,\frac12, 0,\frac12, 0)\rangle - 
|(x,y,0, \frac12,\frac12, 0)\rangle, \\
& |(x,y,\frac12, 0, 0,\frac12)\rangle - 
|(x,y,0, \frac12, 0,\frac12)\rangle, \nn
\eea
where $\pm 1$ denote eigenvalues 
under the action of $\theta$.
Then there are six states of $27\in E_6$, 
but the negative eigenvalue state does not make $\overline{27}$ state 
because it does not cancel the $\Delta$-phase (\ref{phase-orbi}).
In this way we confirm the number of $27$ states is 34, 
and that of $\overline{27}$ states is 0.
The untwisted sector is the same as factorizable model, 
that is $h^{1,1}_{untwisted}=3$ and $h^{2,1}_{untwisted}= 1$.
This is because the untwisted sector is determined by local action of orbifold
and not affected by global structure of Lie lattice. 

This result is confirmed by the Euler number (\ref{euler}),
\bea
\chi &=& \frac{1}{8}\left( 24 \chi_{\theta.\phi}
+\chi_{1.\theta \phi} +\chi_{\theta^2.\phi} \right),\nn \\
\chi_{-1} &=& \frac{1}{8}\left( -24 \chi_{\theta.\phi}
+\chi_{1.\theta \phi} +\chi_{\theta^2.\phi} \right). 
\eea
Here, $\chi_{-1}/2$ is the generation number of the model 
with discrete torsion $\epsilon=-1$. 
The fixed points by the action of $\theta$ and $\phi$ are 
\bea
&(0,0,0,0,0,0),~(1,0,0,0,0,0),~ 
(\frac12,\frac12,\frac12,\frac12,\frac12,\pm \frac12), \\
&(0,0,\frac12,\pm \frac12,0,0),~(\frac12,\frac12,0,0,\frac12,\pm \frac12). \nn
\eea
The fixed points of action of $\theta \phi$ are 
\bea
&(0,0,0,0,0,0),~(1,0,0,0,0,0),~
(\frac12,\frac12,\frac12,\frac12,\frac12,\pm \frac12), \nn\\
&(0,0,\frac12,\pm \frac12,0,0),~(\frac12,\frac12,0,0,\frac12,\pm \frac12),\\
&(\frac12,\frac12,\underline{\pm\frac12,0},0,0),~
(0,0,\underline{\frac12,0},\frac12,\pm\frac12).  \nn
\eea
The fixed points of action of $\theta^2$ and $\phi$ are 
\bea
&(0,0,0,0,0,0),~(1,0,0,0,0,0),~
(\underline{\frac12,0},\underline{\frac12,0},\underline{\pm\frac12,0}), \nn\\
&(\frac12,\pm \frac12,0,0,0,0),~
(0,0,\frac12,\pm \frac12,0,0),~
(0,0,0,0,\frac12,\pm \frac12),~\\
&(\frac12,\frac12,\underline{\pm\frac12,0},0,0),~
(0,0,\underline{\frac12,0},\frac12,\pm\frac12),~
(\frac12,\frac12,\frac12,\frac12,\frac12,\pm \frac12). \nn
\eea
As a result we have 
$\chi_{\theta.\phi}=8$, 
$\chi_{1.\theta \phi}=16$ and  
$\chi_{\theta^2.\phi}=32$. 
The generation numbers in these models are 
\bea
\chi/2 &=& 36 , \nn\\
\chi_{-1}/2 &=& 12,
\eea
This result is different from that of factorizable tori 
SO(4)$\times$SO(4)$\times T^2$, i.e. $\chi/2=60$ and $\chi_{-1}/2=12$. 
All other models on non-factorizable tori are listed 
in Table \ref{tab:z2z4} of Appendix.A.

\subsection{${\mathbb Z}_4 \times {\mathbb Z}_4$ models}
\hspace{.19in}
The basis of
six dimensional tori SO(12) is given by (\ref{so(2n)root}).
In SO(12) root lattice 
the only consistent point group action, 
except its conjugate representation, is
\bea
\theta&=&\left(
\begin{array}{ccccccc}
0 & -1 & 0 & 0 & 0 & 0 \\
1 & 0 & 0 & 0 & 0 & 0 \\
0 & 0 & 0 & 1 & 0 & 0 \\
0 & 0 & -1 & 0 & 0 & 0 \\
0 & 0 & 0 & 0 & 1 & 0 \\
0 & 0 & 0 & 0 & 0 & 1 
\end{array}
\right),\nn \\
\phi&=& \left(
\begin{array}{ccccccc}
1 & 0 & 0 & 0 & 0 & 0 \\
0 & 1 & 0 & 0 & 0 & 0 \\
0 & 0 & 0 & -1 & 0 & 0 \\
0 & 0 & 1 & 0 & 0 & 0 \\
0 & 0 & 0 & 0 & 0 & 1 \\
0 & 0 & 0 & 0 & -1 & 0 
\end{array}
\right).
\label{so(12)z4z4}
\eea

The $\theta^i \phi^j$-twisted sector $T_{ij}$ localizes 
at fixed points or tori as follows, 
\bea
\begin{array}{ll}
T_{01}: & (x,y,0,0,0,0),~(x,y, \frac12, \frac12, \frac12, \frac12), \\
T_{02}: & (x,y,0,0,0,0),~(x,y,\underline{\frac12, \frac12,0,0}),~
(x,y,\frac12, \frac12, \frac12, \frac12), \\
T_{10}: & (0,0,0,0,x,y),~(\frac12, \frac12,\frac12, \frac12,x,y), \\
T_{11}: & (0,0,x,y,0,0),~(\frac12, \frac12,x,y,\frac12, \frac12),\\
T_{12}: & (0,0,0,0,0,0),~(1,0,0,0,0,0),~
(\frac12, \frac12,\frac12, \frac12,\frac12, \pm \frac12), \\
&(\frac12, \frac12, \frac12,\pm \frac12,0,0),~ 
(0,0,0,0, \frac12,\pm \frac12), \\
&(\frac12, \frac12,0,0,\underline{\pm\frac12, 0}),~ 
(0,0,\frac12, \frac12,\underline{\pm\frac12, 0}), \\
T_{13}: & (0,0,0,0,0,0),~(1,0,0,0,0,0),~
(\frac12, \frac12,\frac12, \frac12,\frac12, \pm \frac12), \\
 & (0,0,\frac12, \pm \frac12,0,0),~
(\frac12, \frac12,0,0,\frac12, \pm \frac12), \\
 & (\frac12, \frac12,\underline{\pm \frac12, 0},0,0),~
(0,0,\underline{\frac12, 0},\frac12,\pm \frac12), \\
T_{20}:&(0,0,0,0,x,y),~ (\underline{\frac12,\frac12,0,0},x,y),~
(\frac12, \frac12,\frac12, \frac12,x,y), \\
T_{22}: & (0,0,x,y,0,0),~(\underline{\frac12,\frac12},x,y,\underline{0,0}),~
(\frac12, \frac12,x,y,\frac12,\frac12), \\
T_{23}: & (0,0,0,0,0,0),~(1,0,0,0,0,0),~
(\frac12, \frac12,\frac12, \frac12,\frac12, \pm \frac12), \\
&(0,0,\frac12, \frac12, \frac12,\pm \frac12),~ 
( \frac12,\pm \frac12,0,0,0,0), \\
&(\underline{\frac12, 0},\frac12,\pm \frac12,0,0),~ 
(\underline{\frac12, 0},0,0,\frac12,\pm \frac12). 
\end{array}
\eea
There are 51 orbifold invariant states, 
$h^{1,1}_{twisted}=51$ and $h^{2,1}_{twisted}=0$.
The untwisted sector is the same as factorizable model, 
that is $h^{1,1}_{untwisted}=3$ and $h^{2,1}_{untwisted}= 0$.

This result is confirmed by the Euler number (\ref{euler}),
\bea
\chi &=& \frac{1}{16} \{ 96 \chi_{\theta.\phi} 
+24(\chi_{\theta^2.\phi}+\chi_{\theta.\phi^2}+\chi_{\theta \phi,\phi^2})\nn \\ 
&&+12(\chi_{1.\theta \phi^2} +\chi_{1.\theta \phi^3} +\chi_{1.\theta^2 \phi})
+6\chi_{\theta^2.\phi^2} \},\nn \\
\chi_{-1} &=& \frac{1}{16}\{ -96 \chi_{\theta.\phi} 
+24(\chi_{\theta^2.\phi}+\chi_{\theta.\phi^2}+\chi_{\theta \phi,\phi^2})\nn \\
&&+12(\chi_{1.\theta \phi^2} +\chi_{1.\theta \phi^3} +\chi_{1.\theta^2 \phi})
+6\chi_{\theta^2.\phi^2} \},\\
\chi_{\pm i} &=& \frac{1}{16}\{ 
-24(\chi_{\theta^2.\phi}+\chi_{\theta.\phi^2}+\chi_{\theta \phi,\phi^2})\nn \\
&&+12(\chi_{1.\theta \phi^2} +\chi_{1.\theta \phi^3} +\chi_{1.\theta^2 \phi})
+6\chi_{\theta^2.\phi^2} \}.\nn 
\eea
Here, $\chi_{-1}/2$ and $\chi_{\pm i}/2$ are equal to
generation numbers of the models 
with discrete torsion $\epsilon=-1$ and $\epsilon=\pm i$ respectively. 
We can calculate these quantities in a similar manner with the former section, 
and the result is 
\beq
\chi_{\theta.\phi}=4,~
\chi_{1.\theta \phi}=16,~
\chi_{\theta^2.\phi}=8,~
\chi_{\theta^2.\phi^2}=32,
\eeq
and the generation numbers in these models are obtained as 
\bea
\chi/2 &=& 54 , \nn\\
\chi_{-1}/2 &=& 30, \nn \\
\chi_{\pm i}/2 &=& 6,
\eea
This result is different from that of factorizable tori 
SO(4)$\times$SO(4)$\times$SO(4), i.e. 
$\chi/2=90$, $\chi_{-1}/2=42$ and $\chi_{\pm i}/2=-6$. 

In the ${\mathbb Z}_4 \times {\mathbb Z}_4$ orbifold, 
only three lattices are allowed, 
and the last one is SO(8)$\times$SO(4).
The basis of
six dimensional tori SO(8)$\times$SO(4) are given by
\bea
\alpha_{1} &=&(1,-1,0,0,0,0),\nn \\
\alpha_{2} &=&(0,1,-1,0,0,0),\nn \\
\alpha_{3} &=&(0,0,1,-1,0,0),\nn \\
\alpha_{4} &=&(0,0,1,1,0,0), \\
\alpha_{5} &=&(0,0,0,0,1,-1),\nn \\
\alpha_{6} &=&(0,0,0,0,1,1).\nn 
\eea
The point group actions are 
the same as that of (\ref{so(12)z4z4}). 
However the number of fixed tori is different from 
that of SO(12) lattice.
The numbers of fixed points are 
\beq
\chi_{\theta.\phi}=4,~
\chi_{1.\theta \phi}=16,~
\chi_{\theta^2.\phi}=8,~
\chi_{\theta^2.\phi^2}=32,
\eeq
and the generation numbers of these models are obtained as 
\bea
\chi/2 &=& 60 , \nn\\
\chi_{-1}/2 &=& 36, \nn \\
\chi_{\pm i}/2 &=& 0.
\eea

These are all models which are allowed 
in the ${\mathbb Z}_4 \times {\mathbb Z}_4$ orbifold 
on Lie lattices.

\subsection{${\mathbb Z}_2 \times {\mathbb Z}_3$ models}
\hspace{.19in}
The point group elements $\theta$, $\phi$ 
of ${\mathbb Z}_2 \times {\mathbb Z}_3$ orbifold 
can be expressed by one element 
$\theta \phi \in {\mathbb Z}_6$ which is non-degenerate. 
Let $\omega$ be defined by $\omega \equiv \theta \phi$.
Then we have 
$\theta = \omega^4$ and $\phi = \omega^3$.
This implies ${\mathbb Z}_2 \times {\mathbb Z}_3$ orbifold is 
essentially non-degenerate and
does not provide new models which have different Euler number 
compared to factorizable model.
The Euler characteristic of 
${\mathbb Z}_2 \times {\mathbb Z}_3$ orbifold 
is evaluated by (\ref{euler}),
and it is simplified to
\beq
\chi = 4 \chi_{1.\omega}.\\
\eeq
We see that only non-degenerate element $\omega \in {\mathbb Z}_6$ 
contributes to the generation numbers of these models.
However the Hodge numbers are dependent on lattices as we see below.

For example 
the basis of
six dimensional tori SO(6)$\times$SU(3)$\times$SU(2) is given by
\bea
\alpha_{1} &=&(1,-1,0,0,0,0,0),\nn \\
\alpha_{2} &=&(0,1,-1,0,0,0,0),\nn \\
\alpha_{3} &=&(0,1,1,0,0,0,0),\nn \\
\alpha_{4} &=&(0,0,0,1,-1,0,0), \\
\alpha_{5} &=&(0,0,0,0,1,-1,0),\nn \\
\alpha_{6} &=&(0,0,0,0,0,0,1).\nn 
\eea

\begin{figure}[tbp]
\begin{center}
\includegraphics{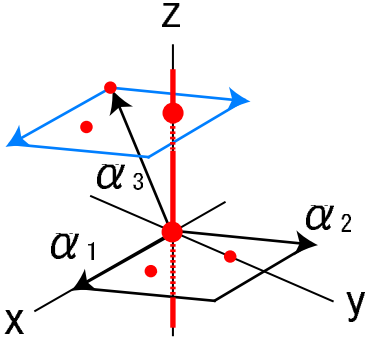}
\caption{${\mathbb Z}_2 \times {\mathbb Z}_3$ 
non-factorizable orbifold.
Circles are points on the $\theta$ fixed tori 
which are parallel to z-axis. 
Red colored line on z-axis is a fixed tori by $\theta$ action, 
and we can see two circles are on the line.
Due to lattice structure all circles are on the same fixed torus. 
So there is only one fixed torus in $\theta$-twisted sector.}
\label{su(3)lattice}
\end{center}
\end{figure}

In the SO(6)$\times$SU(3)$\times$SU(2) root lattice 
the only consistent point group action, 
except its conjugate, is
\bea
\theta&=&\left(
\begin{array}{ccccccc}
0 & 1 & 0 & 0 & 0 & 0 & 0 \\
0 & 0 & 1 & 0 & 0 & 0 & 0 \\
1 & 0 & 0 & 0 & 0 & 0 & 0 \\
0 & 0 & 0 & 0 & 1 & 0 & 0 \\
0 & 0 & 0 & 0 & 0 & 1 & 0 \\
0 & 0 & 0 & 1 & 0 & 0 & 0 \\
0 & 0 & 0 & 0 & 0 & 0 & 1 
\end{array}
\right),\\
\phi&=&
diag(-1,-1,-1,1,1,1,-1).\nn 
\eea
In SO(6) root lattice subspace there is 
only one fixed tori, which is depicted as figure \ref{su(3)lattice}.

The Hodge numbers are calculated as 
$h^{1,1}_{twisted}=26$ and $h^{2,1}_{twisted}=4$ 
from the $\omega^i$-twisted sector $T_{i}$ localizing 
at fixed points or tori.
The untwisted sector is the same as factorizable model, 
that is $h^{1,1}_{untwisted}=3$ and $h^{2,1}_{untwisted}= 1$.
As we have mentioned the number of generations is 24, 
which is the same as the factorizable model. 
However this model has different Hodge numbers from that of
the factorizable model on SU(3)${}^3$ root lattice. 

Similarly the other two 
${\mathbb Z}_2 \times {\mathbb Z}_3$ models 
are examined, 
and the results are listed in Appendix A. 
The non-factorizable models are not same orbifolds 
as factorizable one and 
the structure of Yukawa coupling can be different from 
the factorizable model.

\section{Conclusion and discussion}
\hspace{.19in}
 In this paper we have generalized 
${\mathbb Z}_2 \times {\mathbb Z}_2$ 
non-factorizable orbifold \cite{Faraggi:2006bs}
to ${\mathbb Z}_N \times {\mathbb Z}_M$, 
and show that 
it provides new classes of abelian orbifolds.
In Appendix A,
we give fairly complete classification of 
non-factorizable abelian orbifolds on 
SO($N$) and SU($N$) root lattices. 
The generation numbers of the models $\chi_{\epsilon}/2$ 
are always multiples of 12. 
As we have explained in sections 3 and 4, 
the Euler numbers of non-factorizable orbifolds 
are smaller than that of factorizable models. 
This will be favorable for phenomenological motivation 
to have three generation matter. 
Moreover the Yukawa coupling can be changed in 
non-factorizable models \cite{Forste:2006wq}, 
and there is a possibility to realize the quark and lepton mass matrix with 
appropriate mixing. 

To construct realistic orbifold heterotic string, 
introducing Wilson lines will be important. 
It is interesting to look for realistic models 
using the techniques introduced in this paper. 
There may be correspondence between 
${\mathbb Z}_N \times {\mathbb Z}_M$ non-factorizable orbifolds 
and the free fermionic models of heterotic string. 
We see that there are wide classes of 
${\mathbb Z}_N \times {\mathbb Z}_M$ orbifolds,
and they are interesting in their own right.

\subsection*{Acknowledgement}
\hspace{.19in}
I would like to thank T.Kobayashi for valuable discussions and advice, 
and M.Hanada, and 
the other members of particle physics group of Kyoto University. 
I am grateful to P.Vaudrevange and S.Ramos-Sanchez 
for pointing out some errors on the tables, 
and interesting coincidence \cite{Ploger:2007iq}. 
K.~T.\/ is supported by the Grand-in-Aid for Scientific Research \#172131.

\appendix

\section{Non-factorizable orbifold models}
\hspace{.19in}
In this appendix we list the all non-factorizable models, 
especially the compactified Lie lattices and 
their generation numbers.
All results of our paper are listed here. 
We observe the generation numbers of the models $\chi_{\epsilon}/2$
are always multiples of 12.

General remark is as follows.
If we give the Lie lattice 
for ${\mathbb Z}_N \times {\mathbb Z}_M$ orbifold, 
the point group elements are uniquely determined 
up to their conjugate representations. 
Therefore we classify ${\mathbb Z}_N \times {\mathbb Z}_M$ orbifolds 
by their Lie lattices.
As we include Weyl group and graph automorphisms to the point groups, 
SU(3) and G${}_2$ root lattices are equivalent, and 
the same for SO(8) and F${}_4$ root lattices, 
SO($2N$) and SO($2N+1$) root lattices respectively.

We also calculate the generation numbers $\chi/2$ by
\beq
\frac{\chi}{2} = h^{1,1}_{untwisted} - h^{2,1}_{untwisted} 
+h^{1,1}_{twisted} - h^{2,1}_{twisted}, 
\eeq
where $h^{1,1}_{twisted}$ and $h^{2,1}_{twisted}$ 
are the numbers of chiral matter fields 
from twisted sectors 
with representation in $27$ and $\overline{27}$ respectively.
Similarly $h^{1,1}_{untwisted}$ and $h^{2,1}_{untwisted}$ are 
that of untwisted sectors, 
and they are independent of Lattice structure.

\subsection{${\mathbb Z}_2 \times {\mathbb Z}_2$}
\hspace{.19in}
The result of ${\mathbb Z}_2 \times {\mathbb Z}_2$ orbifold models are 
examined in ref \cite{Faraggi:2006bs, Forste:2006wq}. 
The Euler number is simplified as 
\beq
\chi= \frac{3}{2} \chi_{\theta,\phi}.
\eeq
The allowed values of discrete torsion are 
\beq
\epsilon= \pm1.
\eeq
However these discrete torsions do not make difference 
of generation numbers, except its sign.
For all models, 
the numbers of zero modes of untwisted sector are
\beq
h^{1,1}_{untwisted}= h^{2,1}_{untwisted}= 3.
\eeq
The Hodge number of twisted sectors 
and the generation numbers of 
${\mathbb Z}_2 \times {\mathbb Z}_2$ orbifold models are listed 
in table \ref{tab:z2z2}.
The factorizable model is expressed as $T^2\times T^2\times T^2$, 
because the complex structure of each torus is not fixed by 
orbifold action.
\begin{table}[htbp]
\begin{center}
\begin{tabular}{| c || c| c| c| c|}
\hline 
Lattice & $\epsilon$ &  $ \chi $ & $h^{1,1}_{twisted}$ 
& $h^{2,1}_{twisted}$ \\
\hline \hline
$T^2\times T^2\times T^2$ & 1 & 96 & 48 & 0 \\
 & -1 & -96 & 0 & 48 \\
\hline
SU(3)$\times$SU(2)${}^4$ & 1 & 48 & 28 & 4 \\
 & -1 & -48 & 4 & 28  \\
\hline
SU(4)$\times$SU(2)${}^3$ & 1 & 48 & 24 & 0 \\
 & -1 & -48 & 0 & 24 \\
\hline
SU(3)${}^2 \times$SU(2)${}^2-I$ & 1 & 24 & 18 & 6 \\
 & -1 & -24 & 6 & 18 \\
\hline
SU(3)${}^2 \times$SU(2)${}^2-II$ & 1 & 24 & 16 & 4 \\
 & -1 & -24 & 4 & 16 \\
\hline
SU(4)$\times$SU(3)$\times$SU(2) & 1 & 24 & 14 & 2 \\
 & -1 & -24 & 2 & 14 \\
\hline
SU(4)${}^2$ & 1 & 24 & 12 & 0 \\
 & -1 & -24 & 0 & 12 \\
\hline
SU(3)${}^3$ & 1 & 12 & 9 & 3 \\
 & -1 & -12 & 3 & 9 \\
\hline
\end{tabular}
\end{center}
\caption{
${\mathbb Z}_2 \times {\mathbb Z}_2$ 
orbifold models for standard embedding:
The second column denotes values of the discrete torsion $\epsilon$.
The generation numbers are given by $\chi/2$. 
\label{tab:z2z2}}
\end{table}

\subsection{${\mathbb Z}_2 \times {\mathbb Z}_4$}
\hspace{.19in}
The Euler number and the number of generations with the discrete torsion are 
\bea
\chi &=& \frac{1}{8}\left( 24 \chi_{\theta.\phi}
+\chi_{1.\theta \phi} +\chi_{\theta^2.\phi} \right),\nn \\
\chi_{-1} &=& \frac{1}{8}\left( -24 \chi_{\theta.\phi}
+\chi_{1.\theta \phi} +\chi_{\theta^2.\phi} \right).
\eea
The allowed values of discrete torsion are 
\beq
\epsilon= \pm1.
\eeq
For all models, 
the number of zero modes of untwisted sector are
\beq
h^{1,1}_{untwisted}=3,~ h^{2,1}_{untwisted}= 1.
\eeq
The ${\mathbb Z}_2 \times {\mathbb Z}_4$ orbifold models are listed 
in table \ref{tab:z2z4}.
The factorizable model is expressed as SO(4)$\times$SO(4)$\times T^2$.
\begin{table}[htbp]
\begin{center}
\begin{tabular}{| c || c| c| c| c|}
\hline 
Lattice & $\epsilon$ &  $ \chi $ & $h^{1,1}_{twisted}$ 
& $h^{2,1}_{twisted}$ \\
\hline \hline
SO(4)$\times$SO(4)$\times T^2$ & 1 & 120 & 58 & 0 \\
          & -1 & 24 & 18 & 8 \\
\hline
SO(6)${}^2$& 1 & 48 & 24 & 2 \\
          & -1 & 24 & 14 & 4 \\
\hline
SO(6)$\times$SO(4)$\times$SU(2)& 1 & 72 & 36 & 2 \\
          & -1 & 24 & 16 & 6 \\
\hline
SO(8)$\times T^2$ & 1 & 96 & 48 & 2 \\
       & -1 & 0 & 8 & 10 \\
\hline
SO(8)$\times$SO(4)& 1 & 72 & 36 & 2 \\
          & -1 & 24 & 16 & 6 \\
\hline
SO(10)$\times$SU(2) & 1 & 72 & 34 & 0 \\
       & -1 & 24 & 14 & 4 \\
\hline
SO(12) & 1 & 72 & 34 & 0 \\
       & -1 & 24 & 14 & 4 \\
\hline
\end{tabular}
\end{center}
\caption{${\mathbb Z}_2 \times {\mathbb Z}_4$ 
orbifold models for standard embedding.
\label{tab:z2z4}}
\end{table}

\subsection{${\mathbb Z}_4 \times {\mathbb Z}_4$}
\hspace{.19in}
The Euler number and the numbers of generations with the discrete torsions are 
\bea
\chi &=& \frac{1}{16} \{ 96 \chi_{\theta.\phi} 
+24(\chi_{\theta^2.\phi}+\chi_{\theta.\phi^2}+\chi_{\theta \phi,\phi^2})\nn \\ 
&&+12(\chi_{1.\theta \phi^2} +\chi_{1.\theta \phi^3} +\chi_{1.\theta^2 \phi})
+6\chi_{\theta^2.\phi^2} \},\nn \\
\chi_{-1} &=& \frac{1}{16}\{ -96 \chi_{\theta.\phi} 
+24(\chi_{\theta^2.\phi}+\chi_{\theta.\phi^2}+\chi_{\theta \phi,\phi^2})\nn \\
&&+12(\chi_{1.\theta \phi^2} +\chi_{1.\theta \phi^3} +\chi_{1.\theta^2 \phi})
+6\chi_{\theta^2.\phi^2} \},\\
\chi_{\pm i} &=& \frac{1}{16}\{ 
-24(\chi_{\theta^2.\phi}+\chi_{\theta.\phi^2}+\chi_{\theta \phi,\phi^2})\nn \\
&&+12(\chi_{1.\theta \phi^2} +\chi_{1.\theta \phi^3} +\chi_{1.\theta^2 \phi})
+6\chi_{\theta^2.\phi^2} \}.\nn 
\eea
The allowed values of discrete torsion are 
\beq
\epsilon= \pm1,~ \pm i.
\eeq
For all models, 
the numbers of zero modes of untwisted sector are
\beq
h^{1,1}_{untwisted}=3,~ h^{2,1}_{untwisted}= 0.
\eeq
The ${\mathbb Z}_4 \times {\mathbb Z}_4$ orbifold models are listed 
in table \ref{tab:z4z4}. 
The factorizable model is expressed as SO(4)$\times$SO(4)$\times$SO(4).
\begin{table}[htbp]
\begin{center}
\begin{tabular}{| c || c| c| c| c|}
\hline 
Lattice & $\epsilon$ &  $ \chi $ & $h^{1,1}_{twisted}$ 
& $h^{2,1}_{twisted}$ \\
\hline \hline
SO(4)$\times$SO(4)$\times$SO(4) & 1 & 180 & 87 & 0 \\
          & -1 & 84 & 39 & 0 \\
          & $\pm i$ & -12 & 3 & 12 \\
\hline
SO(8)$\times$SO(4)& 1 & 120 & 58 & 1 \\
          & -1 & 72 & 34 & 1 \\
          & $\pm i$ & 0  & 6 & 9 \\
\hline
SO(12) & 1 & 108 & 51 & 0 \\
       & -1 & 60 & 27 & 0 \\
       & $\pm i$ & 12 & 9 & 6 \\
\hline
\end{tabular}
\end{center}
\caption{${\mathbb Z}_4 \times {\mathbb Z}_4$ 
orbifold models for standard embedding.
\label{tab:z4z4}}
\end{table}

\subsection{${\mathbb Z}_2 \times {\mathbb Z}_3$}
\hspace{.19in}
The Euler number is simplified as 
\beq
\chi = 4 \chi_{1.\omega},\\
\eeq
The discrete torsion is trivial in this case, i.e.
\beq
\epsilon= 1.
\eeq

For all models, 
the numbers of zero modes of untwisted sector are
\beq
h^{1,1}_{untwisted}=3,~ h^{2,1}_{untwisted}= 1.
\eeq
The ${\mathbb Z}_2 \times {\mathbb Z}_3$ orbifold models are listed 
in table \ref{tab:z2z3}.
As we mentioned before, 
we do not distinguish SU(3) root lattice 
from $G_2$ root lattice, 
then the factorizable model can be expressed as SU(3)${}^3$. 
\begin{table}[htbp]
\begin{center}
\begin{tabular}{| c || c| c| c| c|}
\hline 
Lattice & $\epsilon$ &  $ \chi $ & $h^{1,1}_{twisted}$ 
& $h^{2,1}_{twisted}$ \\
\hline \hline
SU(3)${}^3$& 1 & 48 & 32 & 10 \\
\hline
SO(6)$\times$SU(3)$\times$SU(2) & 1 & 48 & 26  & 4 \\
\hline
SO(8)$\times$SU(3) & 1 & 48 & 26 & 4 \\
\hline
SU(5)$\times$SU(3) & 1 & 48 & 26 & 4 \\
\hline
\end{tabular}
\end{center}
\caption{${\mathbb Z}_2 \times {\mathbb Z}_3$ 
orbifold models for standard embedding.
\label{tab:z2z3}}
\end{table}   

The numbers of $h^{1,1}_{twisted}$ and $h^{2,1}_{twisted}$ are same 
in non-factorizable models. 
This implies they are in equivalent class of orbifolds, 
and connected by continuous deformation of geometric moduli.

\section{Lie lattice and Weyl group \label{weyl}}
\hspace{.19in}
The basics of Weyl reflection, Coxeter elements 
and their explicit representations are given in this appendix. 
The point groups considered so far 
\cite{Bailin:1999nk, Kobayashi:1990fx, Kobayashi:1991rp} are generated 
by Coxeter elements from Cater diagrams or 
generalized Coxeter elements which include graph automorphisms. 
We give explicit representations of 
point group elements generated by Weyl reflections and 
graph automorphisms of Lie lattices. 
Besides the generalized Coxeter elements 
this point group contains non-Coxeter elements. 
We utilize these elements for point group of our orbifold models.

\subsection{Weyl reflection and graph automorphism}
\hspace{.19in}
The Weyl group $\cW$ is generated by the following Weyl reflections
\beq
r_{\alpha_k}:\lambda \raw \lambda -2{\alpha_k \cdot \lambda \over \alpha_k \cdot \alpha_k} \alpha_k,
\eeq
where $\alpha_k$ is a simple root of the Lie lattice. 

Convenient basis for simple root of SO($2N$) lattice can be set 
by $N$ elements vectors, 
\bea
\alpha_{1} &=&(1,-1,0,0,\cdots,0),\nn \\
\alpha_{2} &=&(0,1,-1,0,\cdots,0),\nn \\
& \vdots & \nn \\
\alpha_{N-1} &=&(0,\cdots,0,1,-1),\\
\alpha_{N} &=&(0,\cdots,0,1,1).\nn 
\eea

Then the Weyl reflections of these roots are represented by 
$N \times N$ matrices, 
\beq
r_{\alpha_{n}} = 
\bordermatrix{
 &  &   &   & n & n+1 &   &   \cr
 & 1 & 0 & \cdots & 0 & 0 & \cdots & 0 \cr
 & 0 & 1 & \cdots & 0 & 0 & \cdots & 0 \cr
 & \vdots & \vdots & & \vdots & \vdots & & \vdots \cr
 & 0 & 0 & \cdots & 0 & 1 & \cdots & 0 \cr
 & 0 & 0 & \cdots & 1 & 0 & \cdots & 0 \cr
 & \vdots & \vdots &  & \vdots & \vdots & & \vdots \cr
 &  0 & 0 & \cdots & 0 & 0 & \cdots & 1 }
\begin{array}{ l }
 \\ \\ \\ n \\ n+1 \\ \\ \\
\end{array},
\eeq
for $n=1,\cdots,N-1$, and
\beq
r_{\alpha_{N}} = 
\left( \begin{array}{ c c c c c c}
1 & 0 & \cdots & 0 & 0 & 0 \\
0 & 1 & \cdots &  0 & 0 & 0 \\
\vdots & \vdots &  & \vdots & \vdots & \vdots \\
0 & 0 & \cdots & 1 & 0 & 0 \\
0 & 0 & \cdots & 0 & 0 & -1 \\
0 & 0 & \cdots & 0 & -1 & 0 
\end{array} \right).
\eeq
The graph automorphism $g$ of Cartan diagram is represented as 
\beq
g = 
\left( \begin{array}{ c c c c c}
1 & 0 & \cdots & 0 & 0  \\
0 & 1 & \cdots &  0 & 0  \\
\vdots & \vdots &  & \vdots & \vdots \\
0 & 0 & \cdots & 1 & 0 \\
0 & 0 & \cdots & 0 & -1 
\end{array} \right).
\eeq
One can easily find that 
$r_{\alpha_{n}}$ for $n=1,\cdots, N-1$ generate 
permutation group $S_{N}$. 
For example, the product of 
two Weyl reflections which do not commute each other 
makes up $\bZ_3$ element, and the group is $S_3$. 
Moreover if we add the graph automorphism 
$g \in G_{out}$ to Weyl group, 
we can change the sign of any matrix elements of $S_{N}$.
Then the orders of group $\cW$ and $\{ \cW,G_{out} \}$ are 
evaluated as table \ref{weyl-order}.

For SU($N-1$), we take the basis of $N$ elements, 
$\alpha_{n}$, $n=1,\cdots,N-1$, 
and graph automorphism is given by the following $N\times N$ matrix, 
\beq
g = 
\left( \begin{array}{ c c c c c}
0 & 0 & \cdots & 0 & -1 \\
0 & 0 & \cdots & -1 & 0 \\
\vdots & \vdots &   & \vdots & \vdots \\
0 & -1 & \cdots & 0 & 0 \\
-1 & 0 & \cdots & 0 & 0 
\end{array} \right).
\eeq
We can always diagonalize $g$ by the elements of $\cW$, 
and this is negative of identity matrix, 
$-\id_{N} \equiv diag(-1,-1,\cdots,-1)$, 
this means $\{ \cW,G_{out} \}= \{ \cW, -\id_{N} \}$.
Therefore the order of $\{ \cW,G_{out} \}$ is twice as that of $\cW$, 
as in table \ref{weyl-order}.
\begin{table}[htb]
\renewcommand{\arraystretch}{1.5}
\normalsize
\begin{center}
\begin{tabular}{|c||c|c|}
\hline
  & $ \cW $ & $\{ \cW,G_{out} \}$ \\
\hline
\hline
SO($2N$) & $2^{N-1}N!$ & $2^{N}N!$  	\\
\hline
SU($N$) & $N!$ &  $2N!$ \\
\hline
\end{tabular}
\end{center}
\caption{The order of Weyl group and graph automorphism}
\label{weyl-order}
\end{table}

\subsection{Coxeter element}
\hspace{.19in}
The Coxeter element of the Lie lattice 
is defined by product of all simple roots, 
\beq
D_n \equiv r_{\alpha_1} r_{\alpha_2} \cdots r_{\alpha_N}.
\eeq
The other Coxeter elements, 
which are generated by different ordering of product, 
are conjugate to one another, 
and lead to the same class of orbifold.

In general
there are other non-degenerate elements generated by Weyl reflections.
These non-degenerate orbifold can be classified 
by the Carter diagrams \cite{Schellekens:1987ij}. 
The Coxeter elements of SO(8) from Carter diagrams are
\bea
D_4 &=& r_{\alpha_1} r_{\alpha_2} r_{\alpha_3} r_{\alpha_4} =
\left(
\begin{array}{cccc}
0 & 0 & -1 & 0 \\
1 & 0 & 0 & 0 \\
0 & 1 & 0 & 0 \\
0 & 0 & 0 & -1 
\end{array}
\right),~
\\
D_4(a1)&=&r_{\alpha_1}r_{\alpha_2}r_{\alpha_3}r_{\alpha_2+\alpha_3+\alpha_4}=
\left(
\begin{array}{cccc}
0 & -1 & 0 & 0 \\
1 & 0 & 0 & 0 \\
0 & 0 & 0 & 1 \\
0 & 0 & -1 & 0 
\end{array}
\right),~
\eea
where $r_{\alpha_2+\alpha_3+\alpha_4}$ is 
a Weyl reflection of the root 
generated by the sum of simple roots $\alpha_2+\alpha_3+\alpha_4$. 
Then the order of $D_4$ is six, and that of $D_4(a1)$ is four.
\begin{figure}[htbp]
\begin{center}
\includegraphics{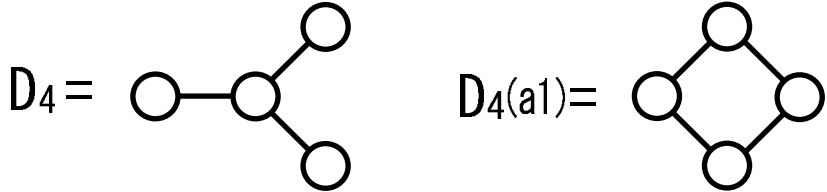}
\caption{Carter diagrams}
\label{so(8)coxeter}
\end{center}
\end{figure}

If we add the graph automorphism of Dynkin diagram to Weyl group, 
we can define generalized Coxeter elements.
For example the SO($2N$) Lie lattice has graph automorphism $g \in G_{out}$ 
which exchanges the simple root $\alpha_{N-1}$ and $\alpha_{N}$, 
then we can define the generalized Coxeter element, 
\beq
C^{[2]} \equiv r_{\alpha_1} r_{\alpha_2} \cdots r_{\alpha_{N-2}} g.
\eeq
The generalized Coxeter element of SO(8) are
\beq
C^{[2]} = r_{\alpha_1} r_{\alpha_2} r_{\alpha_3} g =
\left(
\begin{array}{cccc}
0 & 0 & 0 & -1 \\
1 & 0 & 0 & 0 \\
0 & 1 & 0 & 0 \\
0 & 0 & 1 & 0 
\end{array}
\right).
\eeq
The order of this element is eight.

In the case of SO(8) lattice 
there is another graph automorphism $g^\prime \in G_{out}$, 
which permutes $\alpha_1 \raw \alpha_3 \raw \alpha_4$ cyclically. 
The generalized Coxeter element of this graph automorphism is 
\beq
C^{[3]} \equiv r_{\alpha_1} r_{\alpha_2} g^\prime=
\left(
\begin{array}{cccc}
\frac12 & \frac{-1}{2} & \frac12 & \frac12 \\
\frac12 & \frac12 & \frac12 & \frac{-1}{2} \\
\frac12 & \frac12 & \frac{-1}{2} & \frac12 \\
\frac12 & \frac{-1}{2} & \frac{-1}{2} & \frac{-1}{2} 
\end{array}
\right).
\eeq
However this graph automorphism does not provide new orbifold 
in the non-factorizable models.

\subsection{General point group element on Lie lattice}
\hspace{.19in}
In the case of 
${\mathbb Z}_N \times {\mathbb Z}_M$ orbifolds 
there are other $\cN=1$ supersymmetric models 
which are not generated by Coxeter elements, 
because one element of point group can be degenerate.

We can explicitly see all $\bZ_2$ elements of SO($N$) lattice are
\beq
\theta_{\bZ_2} = 
\left( \begin{array}{ c c c c c c c c c}
0 & 1 &   &   &   &   &   & \cdots & 0\\
1 & 0 &   &   &   &   &   &   & \vdots \\
  &   & \ddots &   &   &   &   &   &  \\
  &   &   & 0 & -1 &   &  &   &   \\
  &   &   & -1 & 0 &   &  &   &   \\
  &   &   &   &   & \ddots &   &   &  \\
  &   &   &   &   &   & -1 &   &  \\
\vdots &   &   &   &   &   &   & \ddots &  \\
0 & \cdots &   &   &   &   &   &   & 1 
\end{array} \right), 
\eeq
and its permutations of Cartesian coordinates. 
The $\bZ_N$ elements are constructed similarly.
For example $\bZ_3$ elements are constructed 
by the following sub-matrices,
\beq
\left(
\begin{array}{ccc}
0 & 0 & 1 \\
1 & 0 & 0 \\
0 & 1 & 0 
\end{array}
\right),~
\left(
\begin{array}{ccc}
0 & 0 & 1 \\
-1 & 0 & 0 \\
0 & -1 & 0 
\end{array}
\right).
\eeq
and their permutations. 
$\bZ_4$ elements includes the following sub-matrices,
\beq
\left(
\begin{array}{cccc}
0 & 0 & 0 & 1 \\
1 & 0 & 0& 0 \\
0 & 1 & 0 & 0 \\
0 & 0 & 1 & 0
\end{array}
\right),~
\left(
\begin{array}{cccc}
0 & 0 & 0 & 1 \\
-1 & 0 & 0 & 0 \\
0 & -1 & 0 & 0 \\
0 & 0 & 1 & 0
\end{array}
\right),~
\left(
\begin{array}{cccc}
0 & 0 & 0 & -1 \\
-1 & 0 & 0 & 0 \\
0 & -1 & 0 & 0 \\
0 & 0 & -1 & 0
\end{array}
\right),~
\left(
\begin{array}{cc}
0 & -1 \\
1 & 0
\end{array}
\right),
\eeq
and their permutations.

In this way we can construct Coxeter and generalized Coxeter elements 
of Weyl group easily and intuitively. 
Similarly all point group elements 
we use in this paper can be expressed 
by Weyl reflections and the graph automorphism.


\begin{thebibliography}{45}

\bibitem{Dixon:1985jw}
  L.~J.~Dixon, J.~A.~Harvey, C.~Vafa and E.~Witten,
  Nucl.\ Phys.\ B {\bf 261} (1985) 678.


\bibitem{Dixon:1986jc}
  L.~J.~Dixon, J.~A.~Harvey, C.~Vafa and E.~Witten,
  Nucl.\ Phys.\ B {\bf 274} (1986) 285.

%
\bibitem{Kobayashi:2004ud}
T.~Kobayashi, S.~Raby and R.~J.~Zhang,
Phys.\ Lett.\ B {\bf 593}, 262 (2004)
[arXiv:hep-ph/0403065];
%
%
Nucl.\ Phys.\ B {\bf 704}, 3 (2005)
  [arXiv:hep-ph/0409098].
%

\bibitem{Forste:2004ie}
S.~F\"orste, H.~P.~Nilles, P.~K.~S.~Vaudrevange and A.~Wingerter,
Phys.\ Rev.\ D {\bf 70}, 106008 (2004);
%
\bibitem{Forste:2005rs}
  S.~F\"orste, H.~P.~Nilles and A.~Wingerter,
  Phys.\ Rev.\ D {\bf 72}, 026001 (2005)
  [arXiv:hep-th/0504117];
%
  Phys.\ Rev.\ D {\bf 73}, 066011 (2006)
  [arXiv:hep-th/0512270].
%
W.~Buchm\"uller, K.~Hamaguchi, O.~Lebedev and M.~Ratz,
 Nucl.\ Phys.\ B {\bf 712}, 139 (2005)
[arXiv:hep-ph/0412318];
%
  Phys.\ Rev.\ Lett.\  {\bf 96}, 121602 (2006)
  [arXiv:hep-ph/0511035];
  %
  arXiv:hep-th/0606187;
%
  H.~P.~Nilles, S.~Ramos-Sanchez, P.~K.~S.~Vaudrevange and A.~Wingerter,
  JHEP {\bf 0604}, 050 (2006)
  [arXiv:hep-th/0603086];


\bibitem{Bailin:1999nk}
  D.~Bailin and A.~Love,
  Phys.\ Rept.\  {\bf 315} (1999) 285.


\bibitem{Faraggi:2006bs}
  A.~E.~Faraggi, S.~Forste and C.~Timirgaziu,
  JHEP {\bf 0608} (2006) 057
  [arXiv:hep-th/0605117].

\bibitem{Forste:2006wq}
  S.~Forste, T.~Kobayashi, H.~Ohki and K.~j.~Takahashi,
  arXiv:hep-th/0612044.


\bibitem{Vafa:1986wx}
  C.~Vafa,
  Nucl.\ Phys.\ B {\bf 273} (1986) 592.


\bibitem{Font:1988mk}
  A.~Font, L.~E.~Ibanez and F.~Quevedo,
  Phys.\ Lett.\ B {\bf 217} (1989) 272.


\bibitem{Kobayashi:1990fx}
  T.~Kobayashi and N.~Ohtsubo,
  Phys.\ Lett.\ B {\bf 262} (1991) 425.

\bibitem{Kobayashi:1991rp}
  T.~Kobayashi and N.~Ohtsubo,
  Int.\ J.\ Mod.\ Phys.\ A {\bf 9} (1994) 87.



\bibitem{Schellekens:1987ij}
  A.~N.~Schellekens and N.~P.~Warner,
  Nucl.\ Phys.\ B {\bf 308} (1988) 397.


\bibitem{Ploger:2007iq}
  F.~Ploger, S.~Ramos-Sanchez, M.~Ratz and P.~K.~S.~Vaudrevange,
  arXiv:hep-th/0702176.



\end{thebibliography}
\end{document}